\documentclass[prb,aps,twocolumn,amsmath,amsmath,amssymb,superscriptaddress,longbibliography]{revtex4-2}

\usepackage{graphicx}
\usepackage{dcolumn}
\usepackage{bm}
\usepackage[colorlinks,linkcolor=blue,citecolor=blue,urlcolor=blue]{hyperref}
\usepackage{float}
\usepackage[dvipsnames]{xcolor}
\usepackage[utf8]{inputenc} 
\usepackage[T1]{fontenc}
\usepackage[normalem]{ulem}
\usepackage{amsmath}

\begin{document}






\title{Phonon-induced pseudogap phase in TiSe$_2$}

\author{Sotirios Fragkos}
\affiliation{Universit\'e de Bordeaux - CNRS - CEA, CELIA, UMR5107, F33405 Talence, France}

\author{Nina Girotto Erhardt}
\affiliation{European Theoretical Spectroscopy Facility, Institute of Condensed Matter and Nanosciences, Université catholique de Louvain, Louvain-la-Neuve, Belgium}
\affiliation{Centre for Advanced Laser Techniques, Institute of Physics, 10000 Zagreb, Croatia}

\author{Evgenia Symeonidou}
\affiliation{Institute of Nanoscience and Nanotechnology, National Center for Scientific Research ‘Demokritos’, 15310 Athens, Greece}

\author{Hibiki Orio}
\affiliation{Experimentelle Physik VII and Würzburg-Dresden Cluster of Excellence ct.qmat, Universität Würzburg, D-97074, Würzburg, Germany}
\affiliation{Ruprecht Haensel Laboratory, Deutsches Elektronen-Synchrotron DESY, D-22607, Hamburg, Germany}
\affiliation{Institute of Experimental and Applied Physics, Kiel University, D-24098, Kiel, Germany}

\author{Dominique Descamps}
\affiliation{Universit\'e de Bordeaux - CNRS - CEA, CELIA, UMR5107, F33405 Talence, France}

\author{Stéphane Petit}
\affiliation{Universit\'e de Bordeaux - CNRS - CEA, CELIA, UMR5107, F33405 Talence, France}

\author{Polychronis Tsipas}
\affiliation{Institute of Nanoscience and Nanotechnology, National Center for Scientific Research ‘Demokritos’, 15310 Athens, Greece}

\author{Kai Rossnagel}
\affiliation{Ruprecht Haensel Laboratory, Deutsches Elektronen-Synchrotron DESY, D-22607, Hamburg, Germany}
\affiliation{Institute of Experimental and Applied Physics, Kiel University, D-24098, Kiel, Germany}

\author{Jakub Schusser}
\affiliation{New Technologies-Research Center, University of West Bohemia in Pilsen, 30614, Pilsen, Czech Republic}

\author{Athanasios Dimoulas}
\affiliation{Institute of Nanoscience and Nanotechnology, National Center for Scientific Research ‘Demokritos’, 15310 Athens, Greece}

\author{Samuel Beaulieu}
\email{samuel.beaulieu@u-bordeaux.fr}
\affiliation{Universit\'e de Bordeaux - CNRS - CEA, CELIA, UMR5107, F33405 Talence, France}

\author{Dino Novko}
\email{dino.novko@gmail.com}
\affiliation{Centre for Advanced Laser Techniques, Institute of Physics, 10000 Zagreb, Croatia}

\begin{abstract}
To comprehend quantum ordered states, such as charge density waves (CDW), in layered transition metal dichalcogenides (TMDCs), it is essential to uncover their underlying normal states. Here, we use time- and angle-resolved extreme ultraviolet photoemission spectroscopy and ab initio electron-phonon calculations to perform excited state band mapping of three prototypical 1T TMDCs, i.e., TiSe$_2$, HfTe$_2$, and ZrTe$_2$, at room temperature. The results reveal the profound impact of strong electron–phonon-induced thermal fluctuations on the normal-phase electronic structure. Specifically, in the moderate electron–phonon coupling regime, as in HfTe$_2$ and ZrTe$_2$, thermal fluctuations only lead to small spectral broadening and band renormalization. In the strongly coupled case, exemplified by TiSe$_2$, we observe soft-phonon-induced, momentum-dependent suppression of spectral weight, i.e., pseudogaps -- extending up to 1~eV above the Fermi level. Our work establishes the normal phase of TiSe$_2$ as a phonon-induced pseudogap phase governed by strong CDW fluctuations, thereby uncovering previously missing aspects of the TiSe$_2$ phase diagram, with broader implications for other TMDCs in the strong electron-phonon coupling regime.
\end{abstract}

\maketitle

Transition metal dichalcogenides with octahedral coordination of the transition metal (1T-TMDCs) are emerging as a fruitful research ground on various exciting ordered states such as superconductivity\,\cite{morosan2006superconductivity}, Mott insulator\,\cite{sipos2008,nakata2021}, chiral order\,\cite{ishioka2010chiral,ren2023}, and charge density waves (CDW)\,\cite{di1976electronic,Rossnagel_2011}, and are often considered as promising candidates for the excitonic condensation at high temperatures\,\cite{cercellier2007evidence,kogar2017sign,gao2023,song2023,gao2024}. As a prominent example, TiSe$_2$ hosts a well-defined strong-coupling CDW in both bulk and monolayer samples\,\cite{di1976electronic,chen2015}, and it is superconducting under doping and pressure\,\cite{morosan2006superconductivity,kusmartseva2009}. On the other hand, CDW signatures in ZrTe$_2$ and HfTe$_2$ were only observed for monolayers\,\cite{yang2022,gao2023,song2023,gao2024}, where the CDW gap and transition temperature $T_{\rm CDW}$ are lower than in TiSe$_2$, pointing to a weaker CDW state. Another curious difference between these 1T-TMDCs is that ZrTe$_2$ and HfTe$_2$ are semimetallic in the normal state\,\cite{aminalragia2017, muhammad2020,Fragkos2021-kz,ren2022,fragkos2024,klipstein1986,nakata2019,youbi2020,tsipas2021}, while the semimetallic or semiconducting nature of TiSe$_2$ is still debated\,\cite{Rossnagel_2011,cercellier2007evidence,li2007,rasch2008}. Although it has been argued that electron correlations are important for shaping the CDW gap and electronic structure\,\cite{bianco2015,hellgren2017,novko2022,hellgren2021,acharya2021}, electron-phonon coupling (EPC) has increasingly been discussed as a key driving force of the CDW phenomena in TiSe$_2$ as in most other TMDCs\,\cite{pashov2025,benic2026}.

%
%
The nature of the true normal state of TiSe$_2$ at room temperature\,\cite{Rossnagel_2011}, i.e., above $T_{\rm CDW}$, remains controversial. On the one hand, transport\,\cite{di1976electronic,knowles2020} and optical measurements\,\cite{li2007} point to the semimetallic electronic structure in the normal phase with a band overlap of around $-100$\,meV. 
On the other hand, angle-resolved photoemission spectroscopy (ARPES) studies reveal a gapped state in TiSe$_2$ above the transition temperature, with the gap size going up to 150\,meV\,\cite{stoffel1985,rossnagel2002charge,kidd2002electron,watson2019orbital}. However, ARPES probes only the occupied states (below the Fermi level $E_F$) and therefore provides no direct access to the unoccupied electronic dispersion, which may be crucial for elucidating the true normal phase of this system. Several studies have investigated doped or self-doped TiSe$_2$ samples to reveal its nature, but again have reported conflicting conclusions\,\cite{rasch2008,zhao2007,may2011,Jaouen2019}. To resolve this issue, it was proposed that TiSe$_2$ is in a pseudogap phase characterized by strong scatterings at the Fermi surface\,\cite{velebit2016scatt,Jaouen2019} in analogy to the state in quasi-1D Peierls systems above $T_{\rm CDW}$\,\cite{lee1973fluct,mckenzie1995pg}. However, these conclusions were again drawn from static photoemission and infrared conductivity measurements, and thus lack access to key information encoded in the electronic structure of unoccupied states.

A powerful technique to overcome these limitations and to investigate excited states and gain information on full band mapping of quantum materials is to utilize ultrafast laser pulses in combination with angle-resolved photoemission spectroscopy (i.e., trARPES), as it was extensively done for extracting band renormalization and carrier thermalization dynamics\,\cite{boschini2024}. This technique was often used to determine the origin and dynamics of CDW states in TMDCs\,\cite{rohwer2011collapse,monney2016revealing,cheng2022light,huber2022mapping}, or to probe the nature of the photo-induced hidden CDW states\,\cite{duan2021optical,duan2023,huber2024ultrafast}. However, the information on the full energy- and momentum-resolved spectra above the $E_F$, which could resolve the aforesaid controversies on the normal phase in TiSe$_2$ and similar TMDCs, is quite limited\,\cite{brube1984}.

Here, we combine trARPES with first-principles EPC calculations to reveal the previously inaccessible excited state's dispersion of TiSe$_2$, and compare it with related TMDCs, namely ZrTe$_2$ and HfTe$_2$, in order to elucidate the role of EPC in shaping the electronic band structure. Our results highlight distinct EPC regimes in these TMDCs, with TiSe$_2$ exhibiting particularly strong coupling between metal $d$ and chalcogen $p$ states near the $E_F$, in contrast to the weaker coupling observed in ZrTe$_2$ and HfTe$_2$. This leads to spectacular differences in energy-resolved excited state population lifetimes measured with trARPES. In addition, this hierarchy of coupling strength is directly linked to the strength of the CDW and ultimately to the degree of the EPC band renormalizations in the normal phase. 
We believe that the full band structure characterization of the room-temperature normal state of TiSe$_2$, as presented here, is an essential step to comprehend other enigmatic features such as the anomalous peak in resistivity\,\cite{di1976electronic,watson2019}, the CDW mechanism\,\cite{Rossnagel_2011}, and the emergent superconductivity\,\cite{morosan2006superconductivity,hinlopen2024}. Beyond that, it shows how strong EPC-induced fluctuations can open the electronic gap, a mechanism that could be relevant for other quantum materials showing gapped or pseudogapped states\,\cite{borisenko2008pg,borisenko2009pg,moon2014,chen2023,liu2025}.

In terms of methodology, we use our time-resolved extreme ultraviolet (XUV) momentum microscopy setup~\cite{Comby22, Fragkos2025-ob, tkach24-2} featuring a polarization-tunable femtosecond infrared (1.2 eV, sub-150 $\mu m$, 0.95 mJ/cm$^2$) pump and femtosecond XUV probe (21.6 eV, sub-50 $\mu m$) pulses, which allows for mapping of the occupied and unoccupied states (up to 1.2\,eV above the $E_F$) over the full photoemission horizon. For more information on the experimental setup, material growth, and decapping procedure, see the Supplemental Material (SM)\,\cite{SM}. Ground state electronic structure calculations were done with the plane-wave basis code Quantum ESPRESSO\,\cite{giannozzi2017qe}, where we use a semi-local PBE exchange-correlation functional. Phonon dynamics were simulated with density functional perturbation theory\,\cite{baroni2001dfpt}, while EPC calculations were performed by means of the EPW code\,\cite{lee2023epw}. Electron spectral functions were obtained from the fully dynamical Fan-Migdal self-energy due to EPC\,\cite{giustino2017epc}. More computational details are presented in the SM\,\cite{SM}.

\begin{figure}[t]
\begin{center}
\includegraphics[width=\columnwidth]{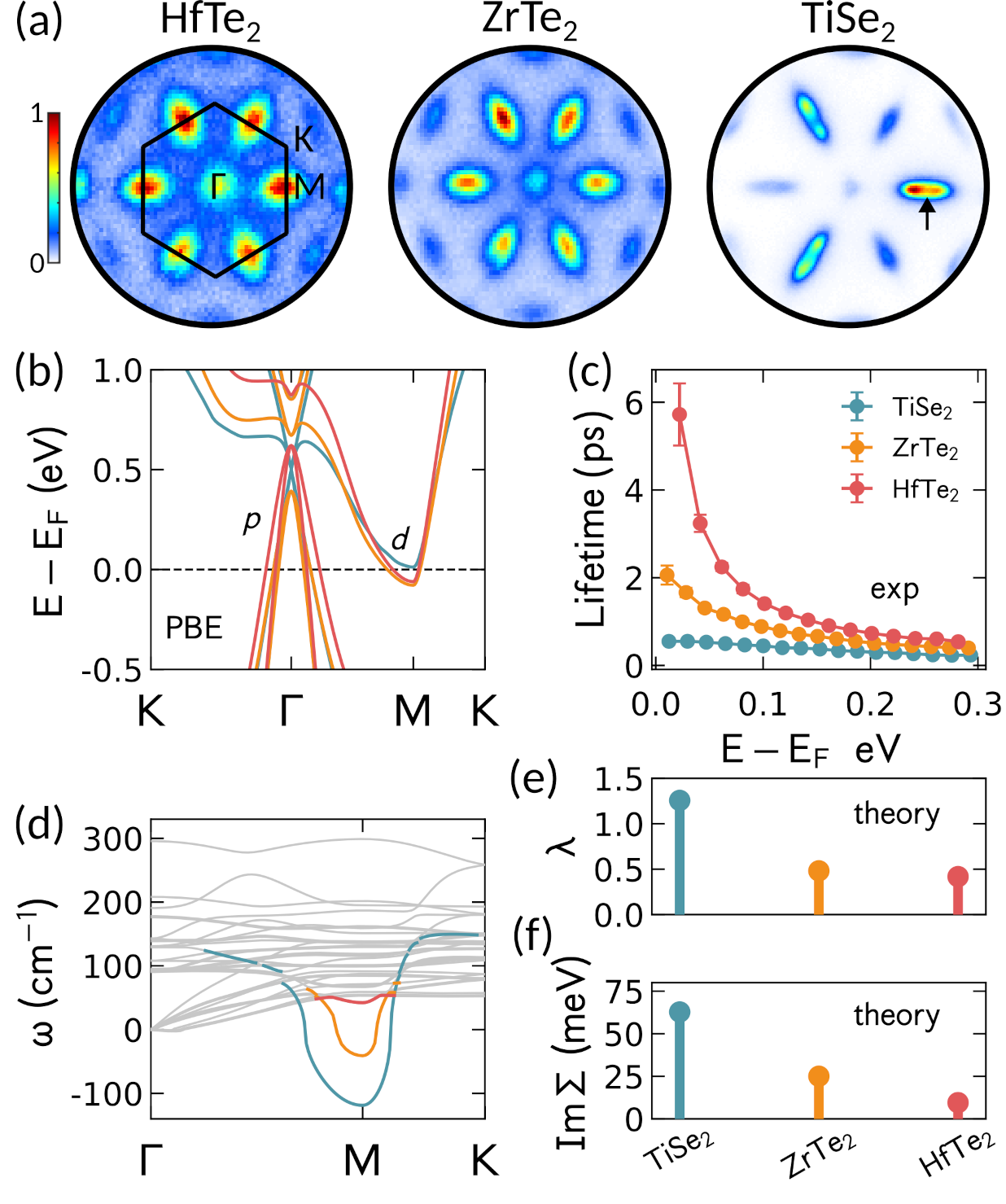}
\caption{\textbf{Excited States Band Mapping of three prototypical 1T-TMDCs.} (a) The constant energy countours (CEC) of excited states at $E_F$ for HfTe$_2$, ZrTe$_2$, and TiSe$_2$ obtained with ultraviolet momentum microscopy (at pump-probe overlap), using s-polarized infrared pump and integrated over all XUV polarization
angles. (b) Theoretical band structure obtained with density functional theory for the three TMDCs. (c) Momentum-integrated photoexcited carrier lifetimes (population thermalization times) as a function of excess energy extracted from the experiments. The ZrTe$_2$ and TiSe$_2$ data are reproduced from our previous publications in Refs.~\cite{fragkos2024,fragkos2026}. (d) Harmonic phonon dispersions at low temperature ($T\approx 0$\,K). The transverse optical phonon Kohn anomalies are highlighted with a different color for each system. (e) The total calculated EPC strengths $\lambda$ and (f) imaginary part of the electron self-energies averaged at $E_F$ for the three systems.}
\label{Fig1}
\end{center}
\end{figure}

We performed trARPES on HfTe$_2$, ZrTe$_2$ and TiSe$_2$ to map the electronic structure over an extended energy-momentum range and to extract energy-resolved population lifetimes.  In Fig.\,\ref{Fig1}(a), we show constant energy contours (CEC) at $E_F$, measured at pump-probe temporal overlap for HfTe$_2$, ZrTe$_2$, and TiSe$_2$, using s-polarized infrared pump and integrated over all XUV polarization angles. In all three cases, we observe electron pockets arising from transition-metal $d$ states at the M points and a hole pocket of the chalcogen $p$ states at the center of the Brillouin zone (BZ). These results for HfTe$_2$ and ZrTe$_2$ are in line with the DFT band structures shown in Fig.\,\ref{Fig1}(b). Namely, the DFT results predict that the hole pocket at the $\Gamma$ point is larger for HfTe$_2$, while the electron pockets at the M point are comparable in size for HfTe$_2$ and ZrTe$_2$ and have an anisotropic elliptical shape. Interestingly, the experimental CEC results at $E_F$ for TiSe$_2$ deviate significantly from the DFT prediction. In particular, the $\Gamma$-point hole pocket in TiSe$_2$ is expected to be of the same size as in ZrTe$_2$, and the M-point electron pocket should show a similar elliptical shape. In contrast, the experimental CEC of TiSe$_2$ shows a small and faint signal at the $\Gamma$ point and a momentum-dependent loss of spectral weight across the middle of the M pockets (see the black arrow). These partially gapped M pockets were so far only observed at the Fermi surface using ARPES measurement at room temperature\,\cite{pillo2000,Jaouen2019,fragkos2026}, and were shown to originate from the strong electron scattering between Ti-$d$ and Se-$p$ pockets\,\cite{yoshiyama1986tise2,fragkos2026} mediated by the soft CDW-related phonon\,\cite{weber2011epc,holt2001x}. The same interpretation was given for the Fermi-surface pseudogaps in NbSe$_2$ and TaSe$_2$\,\cite{borisenko2008pg,borisenko2009pg,kuchinskii2012tmds}, as well as in quasi-1D systems\,\cite{lee1973fluct,mckenzie1995pg}. Here we show that these M-point features remain to be present in the photoexcited conditions. 

Figure \ref{Fig1}(c) shows the experimental momentum-integrated energy-resolved carrier lifetimes for the three studied TMDCs. Similar analysis for ZrTe$_2$ and TiSe$_2$ was done in our previous publications in Refs.~\cite{fragkos2024,fragkos2026}. We observe three different scattering regimes with the lifetimes just above $E_F$ of around 0.6\,ps, 2\,ps, and 6\,ps for TiSe$_2$, ZrTe$_2$, and HfTe$_2$, respectively. We also observe that the carrier thermalization time is independent of momentum, as shown in Fig. S1~\cite{SM}, suggesting that the intervalley ($\Gamma \leftrightarrow \mathrm{M}$) scattering is faster than the temporal resolution of our experimental setup ($\sim 170$\,fs). Assuming that the picosecond dynamics in TMDCs is dominated by electron-phonon scattering\,\cite{caruso2022}, these experimental observations reveal a hierarchy of the EPC strength in these systems, where the strongest (weakest) EPC between chalcogen $p$ and transition metal $d$ states is found in TiSe$_2$ (HfTe$_2$). 

In Figs.\,\ref{Fig1}(d)-(f), we show the results of the \emph{ab initio} EPC calculations to support these conclusions. In particular, we show the low-temperature ($T\approx 0$\,K) harmonic phonon dispersion calculations with highlighted soft phonon modes of the three systems. The Kohn anomaly of the transverse optical phonon mode at $\mathbf{q}=\mathrm{M}$ in TMDCs is ruled by the intervalley EPC between $p$ and $d$ states around the Fermi surface\,\cite{weber2011epc,novko2022,benic2026}. Here, the strongest Kohn anomaly is found in TiSe$_2$, supporting the presence of strong EPC and the low-temperature CDW lattice reconstruction with doubling of the unit cell. On the other hand, the Kohn anomaly in ZrTe$_2$ is much weaker and almost completely suppressed in HfTe$_2$, implying weak EPC regimes. Note also that anharmonic effects, which are not included in the present work, are expected to harden the Kohn anomaly\,\cite{zhou2020anharmonicity,benic2025,benic2026}, which could explain the absence of the CDW in bulk ZrTe$_2$ and HfTe$_2$. Furthermore, in Figs.\,\ref{Fig1}(e)-(f) we show the calculated EPC strengths $\lambda$ and the imaginary part of the electron self-energies due to EPC (i.e., carrier scattering rates) $\mathrm{Im}\,\Sigma$ averaged around the Fermi surface, which are obtained for $T>T_{\rm CDW}$. This confirms the three coupling regimes with $\lambda(\mathrm{TiSe_2})>\lambda(\mathrm{ZrTe_2})>\lambda(\mathrm{HfTe_2})$, where the dominant contribution to the EPC is precisely the intervalley scattering mediated by the soft CDW phonon. See Fig.\,S2 that shows the dominant contribution to the EPC intervalley matrix elements coming from the soft phonon\,\cite{SM}.

\begin{figure}[t]
\begin{center}
\includegraphics[width=\columnwidth]{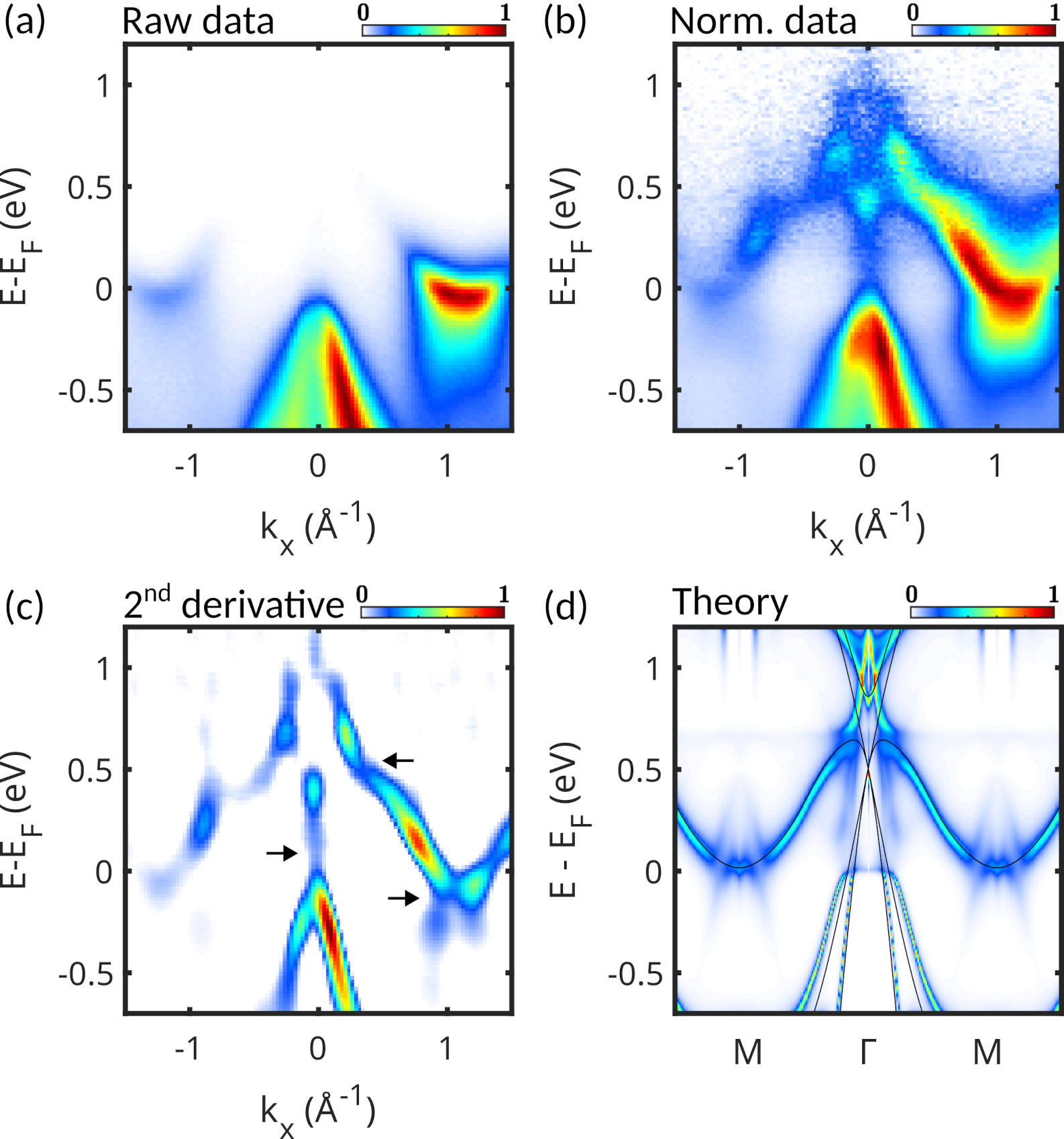}
\caption{\textbf{Experimental and Theoretical Excited States Band Mapping of TiSe$_2$.} (a) Raw, (b) normalized, and (c) second-derivative data for energy-momentum cuts of TiSe$_2$ at pump-probe temporal overlap along the M-$\Gamma$-M$'$ direction, using s-polarized infrared pump and integrated over all XUV polarization
angles. (d) Theoretical electron spectral function along the M-$\Gamma$-M$'$ direction obtained with dynamical EPC self-energy for temperatures just above the transition temperature $T_{\rm CDW}$. The black line shows the DFT band dispersions as obtained with PBE functional.}
\label{Fig2}
\end{center}
\end{figure}

With this analysis, it is clear that the weak EPC in HfTe$_2$ and ZrTe$_2$ does not significantly renormalize the band structure, and the systems remain semimetallic. On the other hand, EPC has a profound impact on TiSe$_2$, where at low temperatures ($T<T_{\rm CDW}$) it induces a CDW reconstruction, while at room temperature ($T>T_{\rm CDW}$) it transforms semimetallic electronic structure into a partially gapped state with strong band renormalizations, which we call a pseudogap phase. In the following, we focus on TiSe$_2$ and analyze in more detail its intriguing excited state band structure in this phase.

\begin{figure*}[ht!]
\begin{center}
\includegraphics[width=0.9\textwidth]{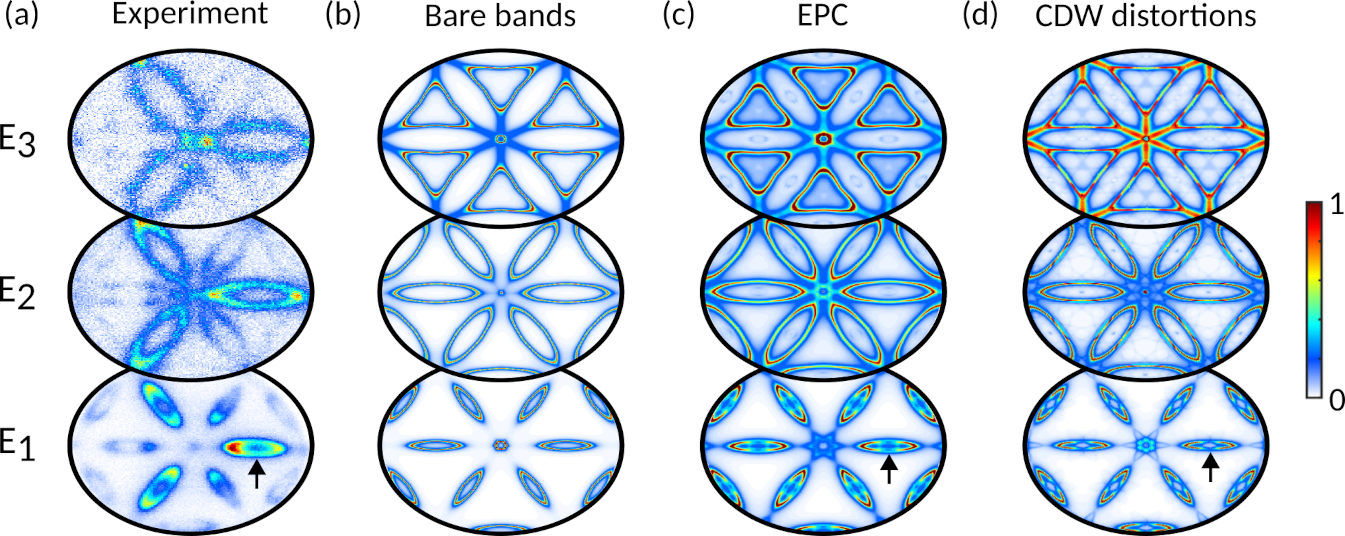}
\caption{\textbf{Full Momentum Band Mapping of TiSe$_2$ of Excited States.} (a) Constant energy contours (CEC) for different binding energies above the $E_F$, obtained by summing photoemission intensities measured with s- and p-polarized light. The values for the energy cuts are $E_1=E_F+200$\,meV, $E_2=E_F+500$\,meV, and $E_3=E_F+750$\,meV. The corresponding theoretical CEC for (b) the bare DFT band structure, (c) electron spectral function obtained with dynamical EPC included, (d) the DFT band structure with small CDW periodic lattice distortions (PLDs) of $\mathrm{\delta Ti=0.057}$\,\AA.
}
\label{Fig3}
\end{center}
\end{figure*}

Precise excited-state band mapping is challenging due to the intrinsically low photoinduced populations above $E_F$ and the short lifetimes of these states. Indeed, the photoinduced signal above $E_F$ is typically orders of magnitude weaker than that from states below it, which are fully occupied in equilibrium, and these excited populations decay on picosecond timescales. To mitigate these limitations and achieve high-statistics band mapping above $E_F$, we stay at the temporal overlap of the IR pump and XUV probe pulses and acquire photoemission spectra over $\sim 24$ hours, with a count rate of a few 10$^5$ electrons per second across the entire BZ. In Figs.\,\ref{Fig2}(a)-(c), we show the nonequilibrium energy-momentum cuts of TiSe$_2$ at pump-probe overlap along the M-$\Gamma$-M$'$ high-symmetry direction (i.e., raw, energy-normalized, and second-derivative data). The results reveal the excited band structure up to around 1.2\,eV above $E_F$, and we consider it to represent the unperturbed electron dispersions of unoccupied states. Besides the well-studied features known from the previous static ARPES data\,\cite{Rossnagel_2011}, i.e., Se-$p$ states at the $\Gamma$ point extending to the $E_F$ and parabolic-like Ti-$d$ band around the M point, we observe several unconventional features (see the black arrows). First of all, we see replicated Se-$p$ bands below the Ti-$d$ band at the M point, which was observed in static ARPES\,\cite{kidd2002electron,cercellier2007evidence}, and recently in the photoexcited TiSe$_2$\,\cite{monney2016revealing,fragkos2026}. These incoherent replica states present at $T>T_{\rm CDW}$ are considered to mark the precursor fluctuating state that precedes the full band folding of Se-$p$ states to the M point once $T<T_{\rm CDW}$. Further, the Ti-$d$ band shows an unexpected kink when approaching the $\Gamma$ point at 0.5\,eV above $E_F$. And surprisingly, we observe a finite incoherent intensity at the $\Gamma$ point from Fermi surface all up to 0.5\,eV. This reveals that the true band structure of TiSe$_2$ at room temperature is not gapped as so far thought\,\cite{rossnagel2002charge,kidd2002electron,rasch2008,cercellier2007evidence,watson2019orbital}.
In contrast, full band mapping of HfTe$_2$ and ZrTe$_2$ reveal semimetallic nature and no signs of unconventional band renormalization (see Fig.\,S3\,\cite{SM}).

Recent trARPES studies of TiSe$_2$ for the CDW phase at $T<T_{\rm CDW}$ have revealed a presence of a transient semimetallic phase, with Se-$p$ and Ti-$d$ band overlap of $\sim 0.4$\,eV, induced by a strong laser excitation ($F\gtrsim 0.1\,\mathrm{mJ/cm^2}$)\,\cite{duan2023,huber2024ultrafast}. Interestingly, the recovery time constants obtained in these trARPES are very similar to the melting dynamics obtained in transient optical spectroscopy\,\cite{porer2014nonthermal}, where they clearly showed from the renormalizations of the collective electron excitation that the CDW is melted to the normal state for fluences above roughly $0.06-0.1\,\mathrm{mJ/cm^2}$. Bearing this connection in mind, it could be that the transient semimetallic phase reported in these trARPES studies is actually the normal phase of TiSe$_2$. Our present results, which uncover that the normal state is a pseudogapped semimetal, support this interpretation.

In Fig.\,\ref{Fig2}(d), we present the theoretical results for the electron spectral function that includes the EPC via the Fan-Migdal self-energy of TiSe$_2$, in comparison with the DFT-PBE band structure. The calculations are done for the normal state lattice just above the $T_{\rm CDW}$ where the CDW phonon is softened\,\cite{weber2011epc}. The results show remarkable renormalizations due to EPC that are in line with the experimental observations. Namely, the theoretical results show faint replica states at the M point, an anomaly of the Ti-$d$ band around 0.5\,eV, and a significant loss of spectral weight of the Se-$p$ band around the $\Gamma$ point. In fact, the strong electron-phonon scattering at the Fermi surface renormalizes the Se-$p$ band from linear dispersion to almost inverse parabolic shape, while opening a partial gap above $E_F$. Note that a weaker EPC for HfTe$_2$ and ZrTe$_2$ does not produce strong band renormalizations beyond the broadening of states (see Fig.\,S4 in SM\,\cite{SM}). Together, these experimental and theoretical observations directly reveal TiSe$_2$ as an unconventional system in which soft-phonon-enhanced thermal fluctuations open the electronic gap. A similar mechanism was discussed to be active in quasi-1D Peierls systems above $T_{\rm CDW}$\,\cite{lee1973fluct,mckenzie1995pg}, as well as in high-$T_c$ cuprates for soft magnetic fluctuations\,\cite{sadovskii2001pg}, and could be relevant for unusual band features observed in strained TiTe$_2$ at room temperature\,\cite{fragkos2019}. Furthermore, dynamical quantum renormalization effects and gap opening associated with a low-energy boson-mediated interaction were also discussed for iron-based pnictides\,\cite{ortenzi2009}. Regarding the case of TiSe$_2$, the role of EPC in opening a pseudogap for $T>T_{\rm CDW}$ was already studied with self-energy calculations on a two-band model\,\cite{yoshiyama1986tise2}, but without momentum resolution and never experimentally confirmed. Similarly, recent molecular dynamics (MD) simulations show how lattice thermal fluctuations open a gap in the normal state, while excitonic correlations are not decisive for its electronic structure\,\cite{pashov2025}. Note that, in some cases, spectral weight suppression could emerge from photoemission matrix elements\,\cite{schusser2022}. However, we argue that this is not the case here, since the latter effect has been shown to be modulated for adjacent M valleys\,\cite{pillo1999,pillo2000}, while we observe pseudogaps in all M pockets within the BZ (e.g., see Fig.\,S5\,\cite{SM}).

In the following, we investigate photoelectron spectra in the full momentum space for energies up to 1\,eV above $E_F$. The photoexcited CEC at the pump-probe temporal overlap for three different energies are depicted in Fig.\,\ref{Fig3}(a) (see Figs.\,S5-S7 for more energy cuts and comparison with HfTe$_2$ and ZrTe$_2$\,\cite{SM}). The results show that the pseudogap openings at the Ti-$d$ pockets along the K-M-K$'$ direction are not only present at $E_F$\,\cite{Jaouen2019,fragkos2026}, but extend all the way to $\sim 0.75$\,eV. Interestingly, this corresponds approximately to the topmost energy of the Se-$p$ state in the bare DFT band structure. For the same energy range, the faint signal is observed at the $\Gamma$ point, as already discussed in Fig.\,\ref{Fig2}. 
Curiously, note that similar pseudogaps at the M point pocket were observed in static ARPES for 1T-TaS$_2$\,\cite{pillo1999}, where Ta-$d$ states are more filled, and the corresponding pocket is comparable in size to Ti-$d$ pocket at higher binding energies\,\cite{Rossnagel_2011}. But the corresponding mechanism of pseudogap creation is assumed to be connected to the electronic origin, i.e., to the hybridization of the overlapping tails of the two Hubbard subbands just above the Mott transition temperature.

In the rest of the panels of Fig.\,\ref{Fig3}, we show the theoretical results at the same energy cuts that include the bare DFT band structure [Fig.\,\ref{Fig3}(b)], spectral function with dynamical EPC included [Fig.\,\ref{Fig3}(c)], and band structure with small CDW periodic lattice distortions (PLDs) [Fig.\,\ref{Fig3}(d)]. The latter calculations are done to account for the PLDs that might still be present in the CDW fluctuating phase above $T>T_{\rm CDW}$. As in Fig.\,\ref{Fig2}(d), the energy cuts with EPC are done for the normal phase TiSe$_2$ structure without any PLDs. Both calculations with lattice contributions [Figs.\,\ref{Fig3}(c)-(d)] show that the suppression (i.e., pseudogaps) of the Ti-$d$ pockets appears at the momenta where these 3 non-equivalent pockets overlap with the Se-$p$ pocket at the $\Gamma$ point in the process of backfolding. In theory, the backfolding in one case is achieved by imposing the small PLDs with $2\times2$ periodicity, and in the other case with thermally-active and strongly-coupled soft phonon with momenta around $\mathbf{q}= \mathrm{M}$ (see Fig.\,S2\,\cite{SM}). Obviously, the bare DFT results do not include the backfolding and thus no pseudogaps are visible in Fig.\,\ref{Fig3}(b).
Note, however, that the results for the small PLDs do not account for the full suppression of the Se-$p$ states at $\Gamma$ for $E \geq E_F$, further emphasizing the importance of the inclusion of the EPC-induced thermal fluctuations via the dynamical self-energy [Fig.\,\ref{Fig3}(c)], which includes the fluctuations of the whole CDW soft phonon mode, as well as other modes with finite EPC, and not just strictly the $\mathrm{q=M}$ part, as it is the case for the small PLDs [Fig.\,\ref{Fig3}(d)]. With this, we show that the gap opening above the Se-$p$ states at the $\Gamma$ point and pseudogaps at the Ti-$d$ pockets around the M point have the same origin, i.e., soft-phonon-mediated intervalley scattering. 

Here we note that the debate on the true nature of the normal phase is directly linked to the debate on the electronic or phonon origin of the CDW in TiSe$_2$\,\cite{Rossnagel_2011,li2007,rasch2008,cercellier2007evidence}. Namely, the semiconducting normal state favors the electronic insulator scenario, since soft exciton is more likely to be stable in a gapped and less screened environment, while the semimetallic normal phase would point to the EPC mechanism, considering that the finite ungapped Fermi surface is a prerequisite for the temperature-dependent Kohn anomaly. Thus, our discovery of the pseudogap state at the Fermi surface and above favors the EPC-dominated mechanism of the CDW transition.


In conclusion, we have performed a detailed excited states band mapping for three prototypical 1T-TMDCs and analyzed the role of EPC in shaping their normal phase electronic band structure. While HfTe$_2$ and ZrTe$_2$ show clear semimetallic behavior with moderate electron-phonon scattering rates, we reveal that TiSe$_2$ has a genuinely different electronic structure due to strong EPC. Namely, strong coupling in TiSe$_2$ leads to the well-known CDW at low temperatures, while above $T_{\rm CDW}$ it induces CDW fluctuations and shapes the electron structure by inducing pseudogaps in both electron Ti-$d$  and hole Se-$p$ pockets, all up to 1\,eV above the $E_F$. Our study identifies electron–phonon thermal fluctuations as an essential microscopic mechanism for understanding charge order, CDW fluctuations, and the electronic structure of TiSe$_2$ above $T_{\rm CDW}$, and more broadly of CDW-bearing TMDCs, with potential implications on emergent superconductivity. 

\section*{Data Availability}

The data supporting the findings of this article will be made openly available on Zenodo upon publication.

\begin{acknowledgments}
We are grateful to Sarath Sasi and Muthu P. T. Masilamani for their assistance on the TiSe$_2$ measurements. We thank Ján Minár, Friedrich Reinert, Maximilian Ünzelmann, and Yann Mairesse for useful discussions. We thank Nikita Fedorov, Romain Delos, Pierre Hericourt, Rodrigue Bouillaud, Laurent Merzeau, and Frank Blais for technical assistance.  We thank Baptiste Fabre for implementing and maintaining the data binning code. 
N.G.E., and D.N. acknowledge financial support from the Croatian Science Foundation (Grants no. IP-2025-02-5926 and UIP-2025-02-5952), European Regional Development Fund for the project ‘Materials for clean energy, advanced sensors and quantum technologies’ (Grant No. PK.1.1.10.0002), and from the project "Podizanje znanstvene izvrsnosti Centra za napredne laserske tehnike (CALTboost)" financed by the European Union through the National Recovery and Resilience Plan 2021-2026 (NRPP). Computational resources have been provided by the Consortium des Équipements de Calcul Intensif (CÉCI), funded by the Fonds de la Recherche Scientifique de Belgique (F.R.S.-FNRS) under Grant No. 2.5020.11 and by the Walloon Region.
The research leading to these results has received funding from LASERLAB-EUROPE (grant agreement no. 871124, European Union’s Horizon 2020 research and innovation programme). We acknowledge the financial support of the IdEx University of Bordeaux / Grand Research Program "GPR LIGHT". We acknowledge support from ERC Starting Grant ERC-2022-STG No.101076639, Quantum Matter Bordeaux, AAP CNRS Tremplin and AAP SMR from Université de Bordeaux. This work is part of the ULTRAFAST and TORNADO projects of PEPR LUMA and was supported by the French National Research Agency, as a part of the France 2030 program, under grants ANR-23-EXLU-0002 and ANR-23-EXLU-0004.  
S.F. acknowledges funding from the European Union’s Horizon Europe research and innovation programme under the Marie Skłodowska-Curie 2024 Postdoctoral Fellowship No 101198277 (TopQMat).
J.S. would like to thank the QM4ST project with Reg. No. \texttt{CZ.02.01.01/00/22\_008/0004572}, cofunded by the ERDF as part of the M\v{S}MT, as well as funding from the European Union’s Horizon Europe research and innovation programme under the Marie Skłodowska‑Curie grant agreement No 101209345—ART.QM funded by the European Union. We acknowledge the funding from MSCA-ITN project SMART-X-860553. 
Views and opinions expressed are however those of the author(s) only and do not necessarily reflect those of the European Union or European Research Executive Agency. Neither the European Union nor the granting authority can be held responsible for them.

\end{acknowledgments}

\bibliography{TiSe2_ExcitedBand.bib} 



\end{document}